\documentstyle[twoside,fleqn,espcrc2]{article}

\def\beq{\begin{equation}}
\def\eeq{\end{equation}}
\def\beqa{\begin{eqnarray}}
\def\eeqa{\end{eqnarray}}
\def\be{\begin{equation}}
\def\ee{\end{equation}}
\def\bea{\begin{eqnarray}}
\def\eea{\end{eqnarray}}

\newcommand{\AmS}{{\protect\the\textfont2
  A\kern-.1667em\lower.5ex\hbox{M}\kern-.125emS}}

\title{The Influence of Dimension Eight Operators on Weak Matrix Elements}

\author{John F. Donoghue\address{Department of Physics, 
        University of Massachusetts, \\ 
        Amherst, MA 01003 USA}%
        \thanks{donoghue@physics.umass.edu. This work supported 
in part by the U.S. National 
Science Foundation}}
       
\begin{document}

\begin{abstract}
I describe recent work with V. Cirigliano and E. Golowich on the effect of
dimension eight operators on weak nonleptonic 
amplitudes. The basic message is that there 
is an inconsistency in the way that many calculations are traditionally
performed. If one calculates matrix elements involving only physics below 
some scale $\mu$, then one needs dimension eight operators explicitly in the
weak OPE. On the other hand if one wants to use dimensional regularization 
throughout, then one needs all scales to be included within the matrix element, 
and this results in the same net effect. A numerical estimate indicates that this
is important below $\mu= 2$~GeV, and this calls into question many of the 
models that have been used to predict $\epsilon'/\epsilon$.
\end{abstract}

\maketitle

\section{Introduction}
In a way, the message of this talk is somewhat surprising as it
implies that we have often not been proceeding correctly in calculating
weak nonleptonic matrix elements over the last 25 years. There needs to be
a significant
modification to standard practice in most models. This is less of an issue
for lattice evaluations but may be fatal for many models that work only
at low scales. 

We normally describe the weak Hamiltonian using the Operator Product
Expansion (OPE)\cite{Wilson1}. In words, we describe this as separating the physics into
that involving energies above a scale $\mu$ and that below. The physics
at high energies is represented by a series of 
local operators and one calculates
the coefficients of these operators using perturbation theory. 
The physics below the
scale $\mu$ goes into the hadronic matrix elements of the operators. 
The OPE for weak matrix elements then reads
\beq
\langle {\cal H}_W^{(\Delta S = 1)} \rangle  \ = \ {G_F \over \sqrt{2}} ~
V_{us}V_{ud}^* \sum_d \sum_{i} ~{\cal C}_i^{(d)} (\mu)~ \langle 
{\cal Q}_i^{(d)} \rangle_\mu \ \ .
\label{dim1}
\eeq 
where $\{ {\cal Q}_i^{(d)} \}$ represents a complete basis set of operators
of increasing dimension $d$, and ${\cal C}_i^{(d)} (\mu)$ are the coefficient 
functions. Dimension-six operators enter the OPE with a dimensionless coefficient, and
dimension-eight would have a coefficient which scales as 1/(mass)$^2$.

Present practice uses dimensional regularization in the calculation of the
coefficient functions\cite{Ciuchini,Buras,overviews}. 
In this case, one has the dimension eight coefficients of
order $1/M_W^2$ and/or $1/m_c^2$~\cite{pivo,Melsom}. For the matrix elements one employs
a framework which describes physics up to the scale $\mu$. It is this combination
that we will show is inconsistent. It must be modified in one of the following ways:
\begin{itemize} 
\item If one calculates the matrix elements including physics only up to the scale $\mu$,
one must include dimension-eight operators in the OPE with coefficient of order
$1/\mu^2$.
\item If instead one wants to use dimensional regularization throughout, one must 
include physics of all scales in the matrix elements. The high energy 
portion of the matrix element above the scale $\mu$ is the effect of dimension-eight
(and higher) operators.
\end{itemize}

\section{Demonstration in a calculable framework}
  In our paper\cite{cdg}, 
we explicitly calculate the dimension eight operators 
within the Standard Model, and we could phrase the argument entirely
within this theory. However, for pedagogical reasons we prefer to 
present the understanding of dimension eight effects using a 
slightly different
interaction. The reason that this is useful is that it builds on 
the theory of QCD sum rules \cite{svz,sumrule}  
in a simple way that lets us 
concretely demonstrate each of the key points. The hamiltonian
which we use has the feature that hadronic matrix elements can
be rigorously related to vacuum polarization functions, which 
are well understood.    

This hamiltonian\cite{dg,others} 
contains one 
left-handed and one right-handed current instead of the usual Standard 
Model hamiltonian in which both currents are left-handed. Specifically
we define\footnote{We define the chiral matrices 
$\Gamma^\mu_{L \atop R} \equiv \gamma^\mu (1 \pm \gamma_5)$.} 
\beqa
{\cal H}_{\rm LR}  &\equiv& 
{g_2^2 \over 8} \int d^4x \ 
{\cal D}_{\mu\nu}(x,M_W^2) ~J^{\mu\nu}(x)  \ \ ,
\nonumber \\
J^{\mu\nu}(x) &\equiv& {1\over 2} 
T \left[ {\bar d}(x) \Gamma^\mu_L u (x) ~
{\bar u}(0) \Gamma^\nu_R s (0) \right]
\nonumber \\
&=& {1\over 2} 
T\left[\big(V^\mu_{1 - i2} (x) + A^\mu_{1 - i2} (x) \big) \right.
\nonumber \\
 & & ~~~~~~\big(V^\nu_{4 + i5}(0) - A^\nu_{4 + i5}(0) \big)\left.\right]   \ ,
\eeqa
where ${\cal D}_{\mu\nu}$ is the $W$-boson propagator and $V^\mu_a$, 
$A^\mu_a \ (a = 1,\dots 8)$ are the flavor-octet vector, 
axialvector currents.

In the chiral limit, the K-to-pi matrix element 
is given in the chiral limit of zero momentum and vanishing 
light-quark masses by the vacuum matrix element
\beqa
{\cal M} &\equiv& \lim_{p=0}{\cal M}(p) 
= {g_2^2 \over 16 F_\pi^2} \int d^4x \ {\cal D}(x,M_W^2) 
\nonumber \\
& &\langle 0 |T\left( V^\mu_3  (x) V_{\mu,3}  (0) - 
A^\mu_3 (x) A_{\mu, 3} (0)\right) | 0 \rangle \ \ .
\label{va3}
\eeqa
Using properties of the vacuum polarization function, this
can be transformed into momentum space
\beqa
{\cal M} &=&  {3 G_F M_W^2 \over 32 \sqrt{2}\pi^2 
F_\pi^2} \int_0^{\infty} dQ^2 \ {Q^4 \over Q^2 + M_W^2} 
\nonumber \\
& & ~~~~~\left[ \Pi_{V,3} (Q^2) - \Pi_{A,3} (Q^2) \right] \ \ .
\label{full}
\eeqa
Here $\Pi_{V,A}$ are the vector and axial vector vacuum polarization 
functions
We will use this simple expression in our analysis. The
reader is interested in more details of how this result
is obtained is referred to the original papers\cite{dg,others}.
However the only essential point for the present argument is
that this matrix element involves an integration over all
scales of some hadronic quantity.

The usual OPE of this matrix element is expressed in terms of
two LR dimension-six
operators. Although it it is easy to perform the 
renormalization group summation, for our purpose here
I just display the OPE to first order in $\alpha_s$,
\beq
{\cal M} \simeq {G_F \over 2 \sqrt{2}F_\pi^2} 
\bigg[ \langle {\cal O}^{(6)}_1 \rangle_{\mu} \ 
+ \ { 3 \over 8 \pi} \ln \left( {M_W^2 \over \mu^2} 
\right) \langle \alpha_s {\cal O}^{(6)}_8 \rangle_{\mu} \bigg] \ \ .
\label{sd10}
\eeq
where
\beqa
{\cal O}^{(6)}_1 &\equiv& {\bar q} \gamma_\mu {\tau_3 \over 2} q
~{\bar q} \gamma^\mu {\tau_3 \over 2} q - 
{\bar q} \gamma_\mu \gamma_5 {\tau_3 \over 2} q 
~{\bar q} \gamma^\mu \gamma_5 {\tau_3 \over 2} q \ \ ,
\nonumber \\
{\cal O}^{(6)}_8 &\equiv& {\bar q} \gamma_\mu \lambda^a
{\tau_3 \over 2} q
~{\bar q} \gamma^\mu \lambda^a {\tau_3 \over 2} q 
\nonumber \\
& & - 
{\bar q} \gamma_\mu \gamma_5 \lambda^a {\tau_3 \over 2} q
~{\bar q} \gamma^\mu \gamma_5 \lambda^a {\tau_3 \over 2} q
\ \ .
\label{sd4}
\eeqa
In the above, $q = u,d,s$, $\tau_3$ is a Pauli (flavor) matrix, 
$\{ \lambda^a \}$ are the Gell~Mann color 
matrices and the subscripts on ${\cal O}^{(6)}_1$, ${\cal O}^{(6)}_8$ 
refer to the color carried by their currents.

Now lets look at the direct calculation of the matrix element using
the vacuum polarization functions.
First we will consider the case where we separate the
physics above and below some value of $Q^2=\mu^2$, with $\mu$ large enough
that we can use perturbation theory for the high energy portion.
Most interesting for our purposes here is the high energy contribution. 
The asymptotic
behavior of the vacuum polarization operator is described by 
the operator product expansion, involving a series of
local operators ordered by increasing dimension. 
In the chiral limit the leading contribution
to the difference of vector and axial-vector correlators is a four-quark 
operator of dimension six\cite{svz,Lanin}, 
followed by a series of higher dimensional 
operators,  
\beq
(\Pi_{V,3} - \Pi_{A,3})(Q^2) \sim 
{2 \pi \langle \alpha_s {\cal O}^{(6)}_8 \rangle_{\mu} \over Q^6} + 
 {{\cal E}^{(8)}_\mu \over Q^8} + \dots \ \ .
\label{c2}
\eeq
Here ${\cal E}^{(8)}_\mu$ represents the combination of
local operators carrying dimension 
eight. These have been discussed and 
partially calculated by Broadhurst and 
Generalis~\cite{bg}. For our purposes, it is not necessary to 
know their specific form, but only the fact of their existence.
Upon performing the integration over $Q^2$ at high energies, 
we find
\beqa
{\cal M}_> (\mu) &=& {3 G_F \over 32 \sqrt{2}\pi^2 F_\pi^2} 
 \left[\right. \ln \left( {M_W^2 \over \mu^2}\right) 2 \pi 
\langle \alpha_s {\cal O}^{(6)}_8 \rangle_{\mu} 
\nonumber \\
& & ~~~~~ +{{\cal E}^{(8)}_\mu\over \mu^2} 
+ \dots \left.\right] \ \ .
\label{dim7}
\eeqa
Here is the first indication of the dimension-eight effect.
In this matrix element it clearly appears as a contribution
to the final answer, it is scaled with
$1/\mu^2$, and it arises from the high energy portion of the calculation,
above $\mu$. 

The low energy portion of the integral goes all into the
matrix element of ${\cal O}^{(6)}_1 $. This is more or less clear, but 
will be shown explicitly below.
The full amplitude is then becomes
\beqa
{\cal M} &\simeq& {G_F \over 2 \sqrt{2}F_\pi^2}
\left[\right.
 \langle {\cal O}^{(6)}_1 \rangle_\mu^{\rm (c.o.)} 
\nonumber \\
& & ~~~+ \ { 3 \over 8 \pi} \ln \left( {M_W^2 \over \mu^2}
\right) \langle \alpha_s {\cal O}^{(6)}_8 \rangle_{\mu} \ 
\nonumber \\
& & ~~~~~  
+{3\over 16 \pi^2}{{\cal E}^{(8)}_\mu \over \mu^2} + \ldots 
\left.\right] \ \ .
\label{sd11}
\eeqa
Comparison of this result with the usual OPE shows that the
dimension-eight term was not properly accounted for in the OPE. 
This is an illustration of the first itemized point in the introduction -
if one performs a strict separation of scales, one needs dimension-eight terms in 
the OPE, scaled by $1/\mu^2$.

However, we can never get $1/\mu^2$ effects in dimensional regularization
because the scale that enters there (which I will call $\mu_{\rm d.r.}$ from now on)
can only appear in logs in 4 dimensions. The point is that dimensional
regularization is not a separation of scales. We can see this explicitly in the 
calculation of the local operator matrix element.
We can obtain this amplitude by taking the $x\to 0$ limit of the
vacuum polarization functions as defined 
in $d$ dimensions, which results in
\beqa
\lefteqn{\langle {\cal O}^{(6)}_1 \rangle_{\mu_{\rm d.r.}}^{\rm (d.r.)} } 
~~~~~~~~~~~~~~~~~~~~~~~~~~~~~~~~~~~~~~~~~~~~~~~~~~~~~~~~~~~~
\nonumber \\
\equiv
\langle 0 |T\left( V^\mu_3  (0) V_{\mu,3}  (0) - 
A^\mu_3 (0) A_{\mu, 3} (0)\right) | 0 \rangle ~~~~& &\\
=  { (d - 1) \mu_{\rm d.r.}^{4 -d} 
\over (4 \pi)^{d/2} \Gamma(d/2)} \int_0^\infty dQ^2 
\ ~Q^d  \left( \Pi_{V,3} - \Pi_{A,3} \right)(Q^2) & &\nonumber 
\label{r2}
\eeqa
When $d < 4$, this expression is finite. 
A key point is that the integral continues to run over
all $Q^2$. Even without evaluating the integral we can
see that to know its value we must include physics from
above the scale $\mu_{\rm d.r.}$, since there is no
separation of scales.

Let us look at the relation of these two schemes
To this end, we split again the $Q^2$ integral into regions below and above 
$Q^2 =\mu^2$. 
For the part of the integration below separation scale $\mu^2$ 
the integral is finite for all dimensions, and we can take the 
limit $d \to 4$. This portion of the integration then 
reproduces exactly the cutoff version of the matrix element. 
The difference between the cutoff and dimensional 
regularization comes entirely from 
the high energy region and again can be calculated
\beqa
\langle {\cal O}^{(6)}_1 
\rangle_{\mu_{\rm d.r.}}^{\rm {\overline {\rm (MS)}}}
&=& \langle {\cal O}^{(6)}_1 \rangle_\mu^{\rm (c.o.)}
\nonumber \\
&+& {3 \alpha_s \over 8 \pi} \left[ \ln\left({\mu_{\rm d.r.}^2 \over 
\mu^2}\right) - {1\over 6}\right] \langle {\cal O}^{(6)}_8 \rangle_\mu 
\nonumber \\
& &+ {3\over 16 \pi^2}{{\cal E}^{(8)}_\mu \over \mu^2} \ \ .
\label{va37}
\eeqa
The mixing with the dimension six operator is expected. (This result
is obtained in the ${\overline {\rm MS}}-NDR$ scheme - see \cite{dg} for
details.) For our purposes, the main point here is that
the high energy region does contribute to the matrix element, through
the dimension eight effect. Comparison with Eq.~(\ref{sd11}) shows that
{\em{all}} of the dimension-eight operator is shifted into the 
${\overline {\rm MS}}$ definition of the dimension-six operator. 
This is seen to be consistent: 
\begin{enumerate}
\item When one performs a separation of scales, one has the need 
for dimension-eight operators in the OPE scaled by
$1 / \mu^2$. 
\item When one defines instead the OPE using dimensional
regularization, one cannot get effects proportional to 
$1 / \mu^2_{\rm d.r.}$, but the same effect appears contained 
within the dimension-six operator matrix element. Overall 
one obtains the same 
total matrix element in either case.\footnote{Since we are treating the 
dimension-six coefficients at leading-log order, we can 
ignore the nonlogarithic dimension-six portion of Eq.~(\ref{va37}). 
To include it only requires an inclusion of the nonlogarithmic 
terms in the coefficient function.}.
\end{enumerate}

Another way to phrase this result is to note that, within dimensional regularization,
the OPE is employed as a 
method for regularizing operators, not for separating scales. 
Regularizaton accounts for the most singular aspects of short distance physics, but
not all of it. The dimension eight effect represents the leading component of the
residual short distance physics. The two ideas of the OPE and the separation of
scales were both put forward by Wilson at about the same time and they have tended to 
be combined together in our thinking. It is interesting that in dimensional 
regularization of nonleptonic operators 
we employ one of Wilson's ideas (the OPE, used a regularization tool)
but not the other.

Within this calculation, we can also calculate reliably the magnitude of 
the dimension eight effect. This is because the vacuum polarization functions
satisfy dispersion relations with the input being given by data on $e^+e^-$ reactions
and $\tau$ decays. The details are found in \cite{cdg,dg}. The relevant comparison 
is between the first and last terms of Eq.~(\ref{va37}), i.e. by how much does the
dimension eight effect shift the matrix element .  
In units of $10^{-7}$ we find
\beq
{\cal M}\  = \  \left\{ 
\begin{array}{ll}
- 0.12 \ + \  0.64 + \dots &  
(\mu = 1~{\rm GeV}) 
\nonumber  \\
- 0.28 \ + \ 0.30 + \dots &
(\mu = 1.5~{\rm GeV}) 
\nonumber  \\
- 0.44  \ + \ 0.17 + \dots &  
(\mu = 2~{\rm GeV}) 
\nonumber  \\
- 0.89 \ + \ 0.04 + \dots &
(\mu = 4~{\rm GeV}) \ \ .
\end{array}
\right.
\label{dim9}
\eeq
where the first entry is the cut-off matrix element and the second is the
dimension eight effect. We see that the effect is of order 100\% 
for $\mu = 1.5$~GeV. At this scale and below, we would also need
to consider yet higher dimension effects, and the whole calculation is
out of control. At higher values of $\mu$, the dimension-eight
effect can be treated as a perturbation on the usual calculation, and it 
is still significant at $\mu = 2$~GeV. 
Certainly at $\mu= 4$~GeV, the effect is small enough to
be neglected.

\section{The problem with standard practice}

The conflict with present practice comes because we mix these two
methods. We use dimensional regularization for the calculation of 
the coefficient functions, but we generally calculate matrix elements 
using a framework that only included physics up to the scale $\mu$.
This is inconsistent and, one way or the other, some physics is missing. 

The present methods for calculating matrix elements involve either lattice 
gauge theory or low energy models. In lattice gauge theory\cite{lattice}, 
there is 
a cutoff of the physics at a momentum scale $p \sim 1/a$, where $a$ is the 
lattice spacing. Most weak matrix element calculations are evaluated at 
$\mu \sim 1/a$, and so are missing the physics beyond this scale.  
In the lattice community there 
already exists the recognition that one needs to add back in the physics 
from higher energies. One example of this insight is embodied in the Syzmansik 
improvement program\cite{symanzik,chris}, 
which does involve adding higher dimension operators. However, it is my understanding that 
this is not yet implemented in weak matrix element calculations.

Low energy models\cite{quarkmodel} 
use quark models or models based on low-energy effective 
field theories. These involve physics which is valid at low energies only.
Therefore these techniques most often involve a cut-off that separates
low-energy from high-energy physics, and includes only low energy physics
in the matrix elements. As we have shown, this is inconsistent when used 
in connection with the usual OPE. Very often the cut-off is taken to be 
very low, $\mu \sim 0.7 \to 1.0$~GeV. This leads to an enormous uncertainty 
due to dimension eight effects. The uncertainty of such methods is then far greater
than was previously estimated.

Most of the estimates of $\epsilon' /\epsilon$ involve low energy models for
the estimate of the gluonic penguin operator matrix element, often labeled
$B_6$. This matrix element has proven difficult to calculate on the lattice, and
therefore in most of the reviews one obtains $B_6$ from low energy models of 
some sort. However, these are all suspect in light of their use of a low cut-off and
their lack of the dimension eight effects. At this stage, I certainly cannot claim
that this explains why the theoretical predictions are below the experimental value, 
as we do not even know the sign of the missing dimension-eight effects. However, I know
of no complete evaluation of $\epsilon' /\epsilon$ that is fully consistent theoretically.

\section{What can be done?}
Let me assume from now on that we agree to calculate using dimensional
regularization. This is the most convenient framework for perturbative
calculations. In this case, there are no dimension eight terms in the 
OPE and we can use all of the past results for the coefficients without
change. However, in this case the matrix elements must be evaluated using 
a method that includes all scales. How can this be done? 

In the case of 
lattice methods, there are several options. One is to simply push the scale
$\mu$ high enough that the residual uncertainty from dimension-eight operators
is small. By our estimates, $\mu \sim 4$~GeV should be sufficient. Alternatively
one may use a range of lattice
spacings and use the data to extrapolate to
the continuum limit at fixed $\mu$. This has been done successfully for the 
kaon B parameter\cite{sharpe}. 
A third option would be to use the Syzmansik improvement
program also for weak matrix elements. This will involve adding in 
dimension-eight operators. 

It will be more difficult to improve many of the quark-model and 
effective field theory models. 
Sometimes these models can be formulated in a way that
includes physics over all scales, even if the short-distance physics 
differs from perturbative QCD (for example, see \cite{peris}).  
These ``all scale'' models can be used
to estimate dimensionally regularized matrix elements if treated
properly. However it is more common in such quark models to use
a cut-off, as one recognizes that the model is not valid at higher
energies. In this case, we would only be able
to add back in the physics at high energies through the use of 
dimension-eight operators. This may be a difficult task. 

The dispersive approach\cite{dg,cg} of my collaborators and myself is a 
special case. Because our approach to two special matrix elements
uses data as input, at least in the chiral
limit, it is not a model. It can be applied over all energy scales and hence
can readily provide dimensionally regularized matrix elements. In our previous 
work the impact of dimension eight operators was seen but not fully understood. It was
treated as an uncertainty in the error bars which were quoted. Now that we have a better
understanding of this effect, we will provide an update to the previous work which
clarifies the proper treatment. 

In our paper\cite{cdg}, we have displayed the dimension eight operators relevant
for the Standard Model at one loop, in a particular method for the
separation of scales. Perhaps these will be useful in estimating the 
magnitude of such effects, or correcting existing calculations.

\end{document}